# Evolution of stellar magnetic activity: probing planet engulfment by red giants

Charlotte Gehan[1]

IRAP, Université de Toulouse, CNRS, CNES, UPS, 14 avenue Edouard Belin, 31400, Toulouse, France ;
charlotte.gehan@irap.omp.eu

October 31, 2025

**ABSTRACT**

It has been observed that the fraction of low-mass (LM) stars ($M \leq 1.5\,M_\odot$) showing photospheric activity in their light curve is larger on the horizontal branch (HB) than during the previous red giant branch (RGB) phase, while the opposite trend has been observed for intermediate-mass (IM) stars ($M > 1.5\,M_\odot$). One hypothesis is that LM red giants (RGs) engulf more planets than IM RGs, which results in a faster surface rotation and a higher magnetic activity. This hypothesis is based on the fact that LM stars reach a maximum radius at the RGB tip that is much larger than for IM stars, making them more likely to engulf planets. However, we need to study the evolution of the active star fraction along the RGB to firmly check this hypothesis. I use independent indicators tracing the activity level in the chromosphere based on the Ca II H&K, H$\alpha$, Mg I and infrared Ca II spectral lines from LAMOST data for $\sim$ 3000 RGs whose evolutionary stage has been identified by asteroseismology with the *Kepler* mission. I find that the fraction of active stars shows different trends for LM and IM stars along the RGB, decreasing for IM stars but unexpectedly increasing for LM stars. Such an increase is not explained by models of single-star evolution and is consistent with the fact that LM stars are more susceptible than IM stars of engulfing planets. Indeed, data shows that IM main-sequence stars exhibit a dearth of planets, consistently with predictions from planet formation theory. In addition, I observe that the fraction of active stars tends to increase for both LM and IM stars on the HB, which stands in partial contrast with previous findings. Finally, I discover that the IM RGB star KIC 9780154 may have engulfed one or more planet(s) as its surface rotation from photometry is twice faster than its envelope rotation from asteroseismology. Characterizing planet engulfment by RGs provides insights into the evolution and fate of most planetary systems, since $\sim$ 97 % of stars pass through the RG evolution stage.

## 1. Introduction

Although the majority of exoplanets are detected around stars on the main sequence (MS), the study of planets orbiting red giants (RGs) provides insights on the impact of stellar evolution on the orbital and physical properties of planetary systems, which is one of the key scientific goals of the future PLATO mission (Rauer et al. 2025). Studying planets around RGs is also crucial to understand the fate of planets around white dwarfs, which represent the ultimate evolutionary stage of $\sim$ 97 % stars. Indeed, the dramatic changes in the stellar structure during the red giant branch (RGB) phase are thought to impact the planets orbits in two opposite ways (e.g., Nordhaus & Spiegel 2013): orbital decay caused by the expansion of the stellar radius, and orbital expansion caused by the stellar mass loss. Characterizing these physical processes requires to observationally study the evolution of planets along the RGB. So far, $\sim$ 440 evolved stars are known to host planets ($\sim$ 230 subgiants and $\sim$ 210 RGs, e.g., Chen et al. 2023).

In the event of a planet being engulfed, the angular momentum conservation of the star-planet system implies an acceleration of the host star's rotation, generating a more efficient dynamo and thus a stronger magnetic field (Aurière et al. 2015). The detection of rapid surface rotation and/or high magnetic activity could thus be a detectable signature of planet engulfment (Privitera et al. 2016a,b; Tayar et al. 2022; Ong et al. 2024). The lack of observed hot Jupiters and the prevalence of multiple planetary systems around subgiants and RGs are potential indicators of planet engulfment as the stellar radius increases (Lillo-Box et al. 2016). Moreover, it is estimated that $\sim$ 10 % of stars with $M \sim 1 - 2\,M_\odot$ will engulf a planet of mass $1 - 10$ Jupiter masses during their evolution as red giants (e.g., O'Connor et al. 2023).

Gaulme et al. (2020) observed that the fraction of active RGs, i.e. exhibiting photometric rotational modulation due to the existence of spots in co-rotation with their photosphere, is larger on the horizontal branch (HB), where stars undergo core-helium burning, than on the RGB, where stars undergo shell-hydrogen burning around the inert helium core, for low-mass (LM) stars with $M \lesssim 1.5\,M_\odot$. The authors interpreted such observation as a potential signature of planet engulfment resulting in LM stars gaining angular momentum during their ascent of the RGB, hence appearing more active once they settle on the HB. This scenario is reinforced by the fact that Gaulme et al. (2020) did not observe such trend for intermediate-mass (IM) stars with $M \gtrsim 1.5\,M_\odot$, for which they found that the fraction of active stars decreases from the RGB to the HB phase as normally expected. This difference in the activity of LM versus IM stars could result from the much smaller maximum radius reached at the RGB tip for IM stars compared to LM stars, allowing LM stars to potentially engulf more planets that can further increase their angular momentum, and as a result, their activity (Gaulme et al. 2020). However, it is not excluded that the activity increase ob-served by Gaulme et al. (2020) for LM stars actually results from an observational bias. Indeed, large and/or long-living spots result in a larger photometric variability than smaller and/or short-living spots, which could lead to a higher detection rate of the associated surface rotation period (Rackham et al. 2018; Basri & Shah 2020). Since the stellar structure changes dramatically between the RGB and the HB, with significant radius expansion then shrinking (Kippenhahn et al. 2013) as well as mass loss (Harper 2018), one could expect the surface magnetic field properties to also be impacted, including the spots distribution and/or properties. It could be that spots are also present for a similar fraction of LM RGB stars than for LM HB stars, but are not detected if smaller and/or shorter-living than on the HB phase.

In order to get a more complete view of the activity of RGs with different evolutionary stages and masses, we need independent indicators tracing the activity level regardless of whether or not spots are detected on the photosphere. To that end, I use several spectral lines whose depth trace the level of chromospheric activity, namely the Ca II H&K lines (3969 & 3934 Å), one of the MgI lines (5184 Å), the H$\alpha$ line (6563 Å) and one of the Ca II infrared (IR) lines (8542 Å). The depth of these lines has been measured by Gehan et al. (2022) and Gehan et al. (2024) using the optical spectra from Data Release 7 of the Large Sky Area Multi-Object Fiber Spectroscopic Telescope survey (LAMOST) for $\sim 3000$ RGs whose photometric variability has been measured previously by Gaulme et al. (2020) using *Kepler* data.

Here, I aim at investigating how the magnetic activity of RGs evolves for LM versus IM stars along the RGB and towards the HB and put my results in perspective with planet engulfment. In Sect. 2, I explain my method to study the measured magnetic activity indicators as a function of stellar evolution and describe the data set I use. I present the results in Sect. 3. In Sect. 4, I discuss the results and emphasizes the observational indications of potential planet engulfment by combining existing envelope rotation measurements, models of internal stellar structure, and properties of the planets known up-to-date. Section 5 is devoted to conclusions.

## 2. Method and data

I use the chromospheric activity indicators measured by Gehan et al. (2024) for $\sim 3000$ RGs using the Ca II H&K lines, the Mg I line at 5184 Å, the H$\alpha$ line and the Ca II IR line at 8542 Å from the optical spectra of Data Release 7 of LAMOST to investigate stellar activity along the RGB and on the HB. The resulting chromospheric indicators are $S_{\rm Ca\,II\,H\&K}$, $S_{\rm Mg\,I}$, $S_{\rm H\alpha}$ and $S_{\rm Ca\,II\,IR}$, respectively.

On the RGB, I consider two sets of stars, one on the early RGB with $\nu_{\rm max} \geq 100\,\mu$Hz and the other slightly more evolved with $\nu_{\rm max} < 100\,\mu$Hz. I use $\nu_{\rm max} = 100\,\mu$Hz to separate the two sets of stars in order to keep a significant number of stars in each set. In addition, I considered separately LM and IM stars on the RGB and on the HB. I considered that LM stars have $M \leq 1.5\,M_\odot$ and IM stars have $M > 1.5\,M_\odot$ in accordance with the findings of Gaulme et al. (2020) that the active star fraction evolves differently between the RGB and the HB for stars below and above $1.5\,M_\odot$, increasing for $M \leq 1.5\,M_\odot$ while decreasing for $M > 1.5\,M_\odot$. The numbers of LM and IM stars for each evolutionary stage and activity indicator are presented in Table 1. $\nu_{\rm max}$ and stellar mass were obtained by Gaulme et al. (2020). I also indicate the corresponding numbers of RGs from Gaulme et al. (2020). In order to take into account the evolution of each activity indicator along the RGB, I fit a power law as a function of $\nu_{\rm max}$ to the stars on the RGB (Fig. 1) in order to get a reference activity level along the RGB, which I also use to compare activity on the HB with respect to the RGB.

For each activity indicator except the H$\alpha$ indicator, I then compute the active star fraction as the fraction of stars with an activity index above the corresponding 2-$\sigma$ prediction interval around the fit. As regards the H$\alpha$ indicator, Gehan et al. (2024) strikingly noticed that it is correlated with the oscillations amplitude, at odds with the expectation that higher activity levels lead to the suppression of oscillations, which was what (Gehan et al. 2024) observed with the other activity indicators. The authors interpreted this observation as resulting from the more important contribution of dark filaments to the H$\alpha$ line compared to the other spectral lines (Meunier & Delfosse 2009). The number and size of dark filaments are expected to increase with the level of activity in the case of the Sun (Mazumder et al. 2021), which could result in a smaller H$\alpha$ indicator at higher stellar activity levels. Furthermore, the absorption in the H$\alpha$ line has been observed to first increase with the activity level in Ca II H&K, before switching to emission for large enough activity in Ca II H&K (Cram & Mullan 1979; Cram & Giampapa 1987; Scandariato et al. 2017; Meunier et al. 2022). This is also consistent with the observation of high H$\alpha$ indices for the RGs in close binaries that also exhibit high activity levels in the other spectral lines (Gehan et al. 2024). Since most of the RGs studied here are not very active, I expect the H$\alpha$ index to be anticorrelated with the activity level on average. Hence, I consider that stars are active in H$\alpha$ if $S_{\rm H\alpha}$ lies below the 2-$\sigma$ prediction interval around the fit, by contrast with the other activity indicators analyzed in this study.

I checked that taking the 3-$\sigma$ prediction interval instead of the 2-$\sigma$ prediction interval around the fit in Fig. 1 does not significantly change the trends derived in this study, although it significantly decreases the number of stars that are considered active (see Appendix A). Indeed, the active star fractions derived using the 3-$\sigma$ prediction interval are much lower and do not agree overall with the active star fractions derived by Gaulme et al. (2020) from the photometric detection of rotation periods. Hence, I keep the 2-$\sigma$ prediction interval in the present study as a threshold to determine the active star fractions from different chromospheric activity indicators.

## 3. Results

The numbers and fractions of active LM and IM stars on the early RGB, medium RGB, and on the HB are presented in Table 2 and displayed in Fig. 2. In addition, I recomputed the fraction of photometrically active RGs from Gaulme et al. (2020) by selecting the RGs that have a detected rotation period and discriminating between LM and IM RGs on the low RGB, the medium RGB, and on the HB in the same way as I did for the other activity indicators.

### 3.1. Fractions of active stars along the red giant branch

On the low RGB, I observe that the fraction of active IM stars is higher than for LM stars for all the activity indicators considered here (Fig. 2). This is expected because during the main sequence, IM stars have a shallow convective envelope, or even no convective envelope at all for $M \geq 1.3\,M_\odot$, contrary to LM stars that have a deep convective envelope. As a consequence, LM stars spin-down due to magnetic braking (Kraft 1967; Skumanich 1972), which consists in a coupling between the surface magnetic field and the stellar wind (Schatzman 1962; Weber & Davis 1967; Mestel 1968). By contrast, IM stars do not undergo

Table 1: Characteristics of the entire sample.

| | RGB, all | | | | HB, all | |
|---|---|---|---|---|---|---|
| | $M \leq 1.5\,M_\odot$, $\nu_{max} \geq 100\,\mu$Hz | $M \leq 1.5\,M_\odot$, $\nu_{max} < 100\,\mu$Hz | $M > 1.5\,M_\odot$, $\nu_{max} \geq 100\,\mu$Hz | $M > 1.5\,M_\odot$, $\nu_{max} < 100\,\mu$Hz | $M \leq 1.5\,M_\odot$ | $M > 1.5\,M_\odot$ |
| Gaulme et al. (2020) | 392 | 401 | 166 | 100 | 1450 | 794 |
| Ca II H&K | 271 | 279 | 111 | 75 | 991 | 508 |
| Mg I | 278 | 281 | 112 | 71 | 1001 | 522 |
| H$\alpha$ | 285 | 295 | 114 | 76 | 1038 | 527 |
| Ca II IR | 282 | 295 | 114 | 77 | 1044 | 529 |

**Notes.** Number of RGB and HB stars with measured activity for two different mass and $\nu_{max}$ intervals; each row corresponds to a given activity indicator.

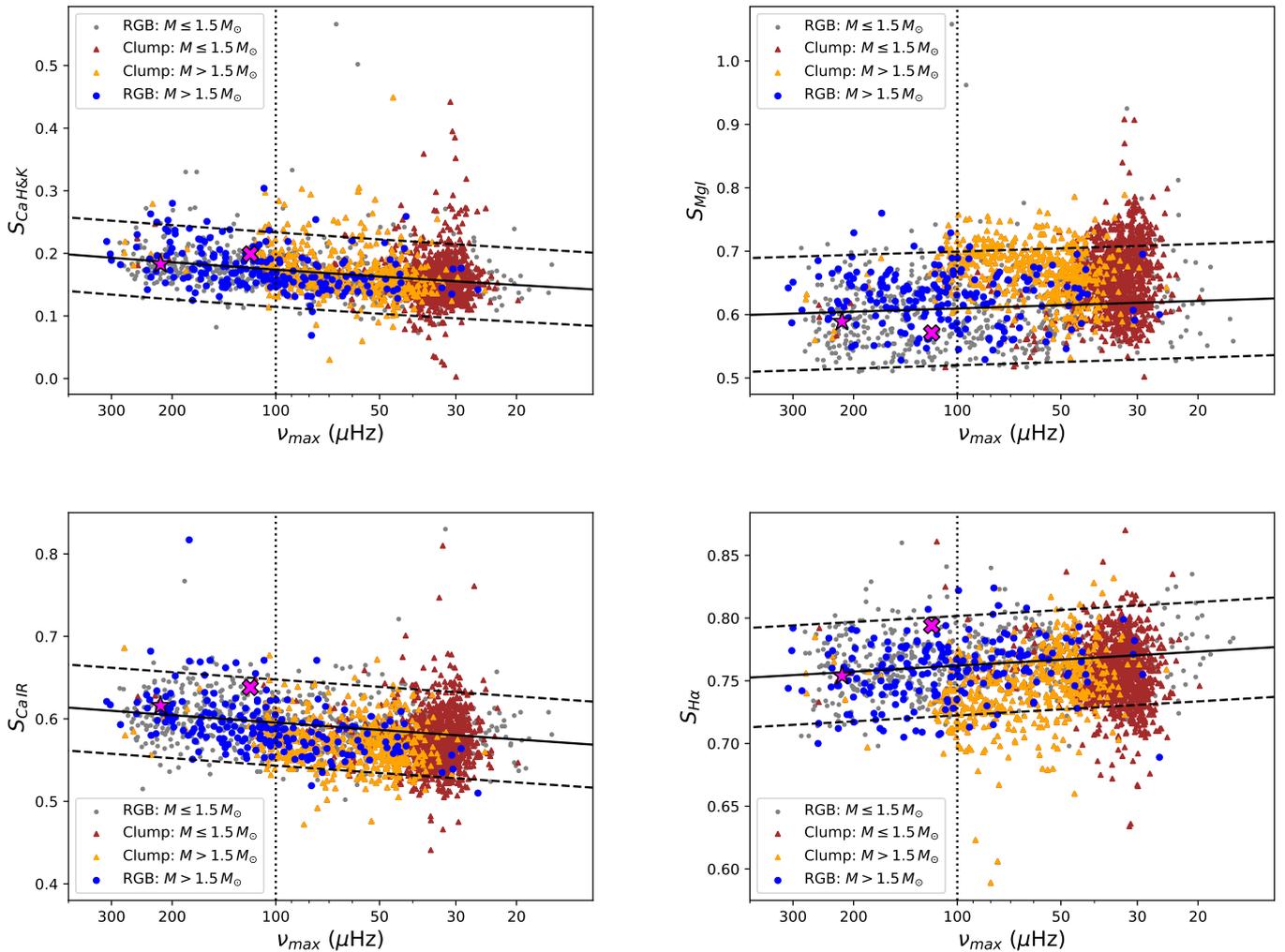

Fig. 1: Activity indicators as a function of $\nu_{max}$ (inverted axis). Low- and intermediate-mass stars on the RGB are represented by grey and blue dots, respectively. Low- and intermediate-mass stars on the HB are represented by red and orange triangles, respectively. KIC 9267654 and KIC 9780154, which may have engulfed one or more planets since their surface rotates faster than their envelope, are shown with the magenta cross and star symbols, respectively. Vertical dotted lines indicate $\nu_{max} = 100\,\mu$Hz. Continuous and dashed lines represent a fit by a power law to the RGB stars and the corresponding 2-$\sigma$ prediction interval around the fit, respectively.

magnetic braking during the main sequence and are therefore expected to enter the RGB with a faster surface rotation and higher activity than LM stars see Sect. 4.1.1.

I observe clear differences in the evolution of the fraction of active LM versus IM stars along the RGB for all the activity indicators used. The fraction of active IM stars decreases for all the activity indicators considered in the present study, while the fraction of active LM stars decreases only for the H$\alpha$ indicator and for RGs with a photometric rotation period as a signature of the existence of magnetic activity, for which the decrease is

Table 2: Characteristics of the active sample.

| | RGB, active | | | | HB, active | |
|---|---|---|---|---|---|---|
| | $M \leq 1.5\,M_\odot$, $\nu_{\mathrm{max}} \geq 100\,\mu\mathrm{Hz}$ | $M \leq 1.5\,M_\odot$, $\nu_{\mathrm{max}} < 100\,\mu\mathrm{Hz}$ | $M > 1.5\,M_\odot$, $\nu_{\mathrm{max}} \geq 100\,\mu\mathrm{Hz}$ | $M > 1.5\,M_\odot$, $\nu_{\mathrm{max}} < 100\,\mu\mathrm{Hz}$ | $M \leq 1.5\,M_\odot$ | $M > 1.5\,M_\odot$ |
| Gaulme et al. (2020) with $P_{\mathrm{rot}}$ | 2.8 % (11) | 0.25 % (1) | 20.5 % (34) | 2.0 % (2) | 3.7 % (53) | 9.1 % (72) |
| Ca II H&K | 3.7 % (10) | 5.0 % (14) | 6.3 % (7) | 2.7 % (2) | 4.6 % (46) | 6.5 % (33) |
| Mg I | 4.0 % (11) | 4.3 % (12) | 4.5 % (5) | 1.4 % (1) | 17.3 % (173) | 19.2 % (100) |
| H$\alpha$ | 4.6 % (13) | 2.7 % (8) | 10.5 % (12) | 5.3 % (4) | 12.4 % (129) | 17.3 % (91) |
| Ca II IR | 4.3 % (12) | 5.1 % (15) | 7.9 % (9) | 1.3 % (1) | 3.9 % (41) | 0.9 % (5) |

**Notes.** Same as Table 1, but showing the percentage of active stars; the number in parenthesis indicates the number of active stars. In the case of chromospheric indicators, stars are considered active if the corresponding activity indicator lies above the 2-$\sigma$ prediction interval around the fit in Fig. 1, except for the H$\alpha$ indicator for wich stars are considered active if $S_{\mathrm{H}\alpha}$ lies below the 2-$\sigma$ prediction interval around the fit.

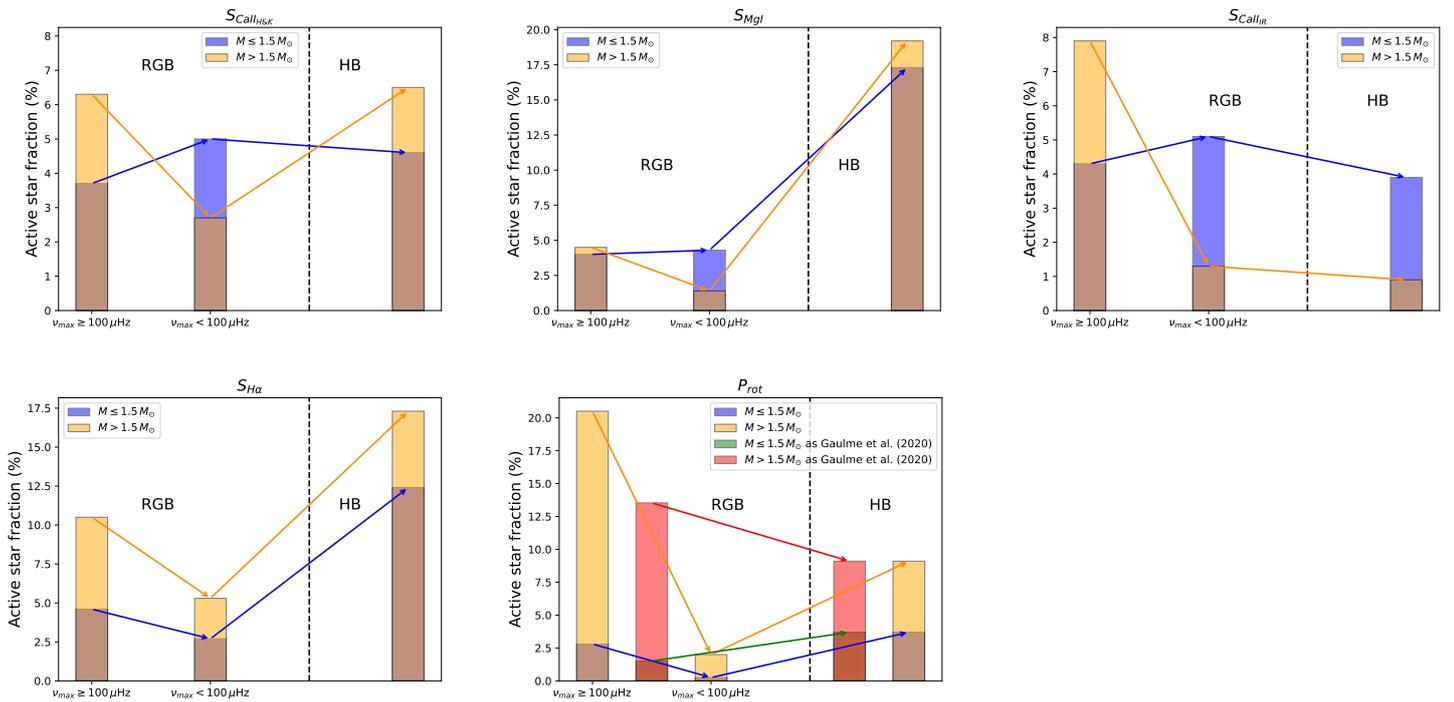

Fig. 2: Evolution of the active star fractions using different activity indicators, from the lower RGB ($\nu_{\mathrm{max}} \geq 100\,\mu\mathrm{Hz}$) to the medium RGB ($\nu_{\mathrm{max}} < 100\,\mu\mathrm{Hz}$) and the HB (on the right-hand side of the vertical dashed lines). Blue bars and arrows correspond to low-mass stars, orange bars and arrows correspond to intermediate-mass stars. The bottom panel indicates in addition the fractions of active LM (green bars and arrows) and IM (red bars and arrows) stars computed over the whole RGB without discriminating between the lower ($\nu_{\mathrm{max}} \geq 100\,\mu\mathrm{Hz}$) and medium ($\nu_{\mathrm{max}} < 100\,\mu\mathrm{Hz}$) RGB.

systematically sharper for IM stars compared to LM stars. I note that the fraction of active LM stars becomes larger than that of IM stars for the Ca II H&K, Mg I and Ca II IR indicators, while it remains below that of active IM stars for the H$\alpha$ indicator and for RGs with a photometric rotation period.

In addition, the left panel of Fig. 3 and Table 3 show that the decrease in the fraction of active stars on the medium RGB compared to the low RGB is larger for IM stars than for LM stars for almost each given activity indicator. The exception lies for RGs with a photometric rotation period as a signature of the existence of magnetic activity, for which I instead observe that the fraction of active stars on the medium RGB decreases more compared to the low RGB for LM stars. Overall, the fractions of active stars between the low and the medium RGB decrease by a median factor of 2.8 for IM stars and increase by a median factor of 1.1 for LM stars when considering only chromospheric activity indicators (see Table 3). This stands in contrast with the results I obtain by by discriminating RGs that have a photometric rotation period on the early and the medium RGB, for which I find an increase of the active stars fraction by a factor of about 10–11 for both LM and IM stars. This may reflect the bias that photometric spot detection can suffer from (Rackham et al. 2018; Basri & Shah 2020).

### 3.2. Fractions of active stars on the horizontal branch

I also observe that the fraction of active stars increases for both LM and IM stars on the HB compared to the medium RGB, for all the activity indicators except the Ca II indicators : the fraction of active LM stars remains roughly the same for the Ca II H&K

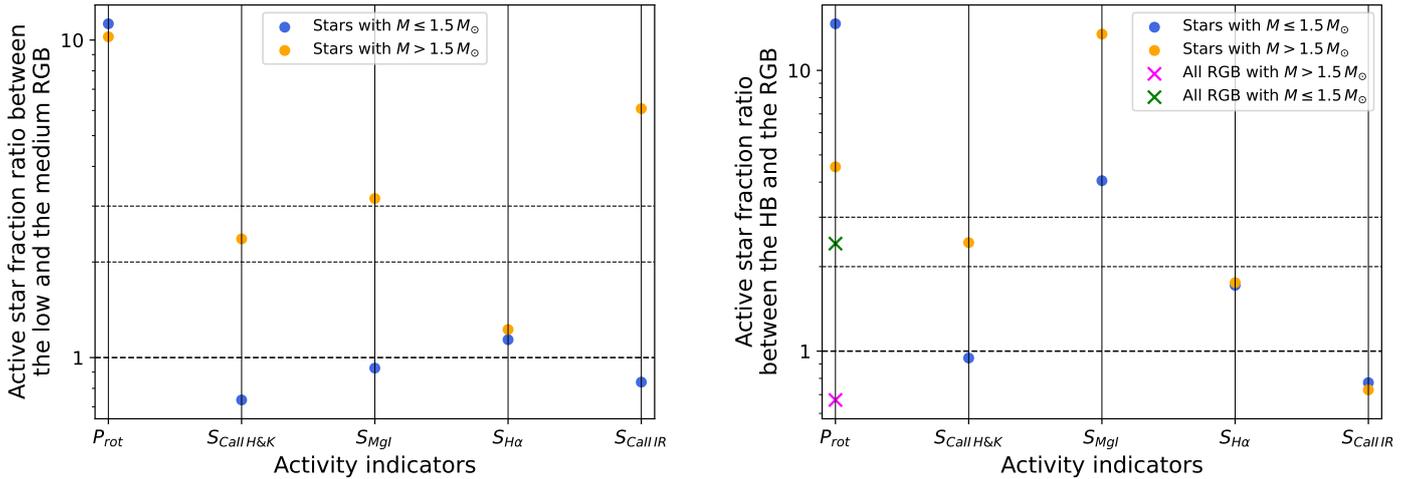

Fig. 3: Ratios between the fractions of active stars for different evolutionary stages for the different activity indicators used in this study (written on the x-axis). LM and IM stars are represented in blue and in orange, respectively. Horizontal dashed lines in bold indicate a ratio of 1, while the other horizontal dashed lines indicate ratios of 2 and 3. *Left:* Ratios between the low RGB ($\nu_{max} \geq 100\,\mu$Hz) and the medium RGB ($\nu_{max} < 100\,\mu$Hz). *Right:* Ratios between the HB and the medium RGB. For RGs with detected $P_{rot}$, we indicate in addition the ratios obtained for active LM (green cross) and IM (magenta cross) stars when considering stars accross over the whole RGB without discriminating between the lower and medium RGB.

Table 3: Ratios between the fractions of active stars on the low and the medium RGB.

| Activity indicator | $M \leq 1.5\,M_\odot$ | $M > 1.5\,M_\odot$ |
|---|---|---|
| $P_{rot}$ | 11.25 | 10.24 |
| $S_{Ca\,II\,H\&K}$ | 0.74 | 2.36 |
| $S_{Mg\,I}$ | 0.93 | 3.17 |
| $S_{H\alpha}$ | 1.68 | 1.83 |
| $S_{Ca\,II\,IR}$ | 0.84 | 6.08 |

**Notes.** Each row corresponds to a given activity indicator. The low RGB corresponds to $\nu_{max} \geq 100\,\mu$Hz, the medium RGB corresponds to $\nu_{max} < 100\,\mu$Hz.

Table 4: Ratios between the fractions of active stars on the HB and the medium RGB.

| Activity indicator | $M \leq 1.5\,M_\odot$ | $M > 1.5\,M_\odot$ |
|---|---|---|
| $P_{rot}$ | 14.66 | 4.53 |
| $S_{Ca\,II\,H\&K}$ | 0.95 | 2.36 |
| $S_{Mg\,I}$ | 4.07 | 13.60 |
| $S_{H\alpha}$ | 4.55 | 3.28 |
| $S_{Ca\,II\,IR}$ | 0.77 | 0.73 |

**Notes.** Each row corresponds to a given activity indicator. The medium RGB corresponds to $\nu_{max} < 100\,\mu$Hz.

Table 5: Properties of KIC 9267654 and KIC 9780154.

| Properties | KIC 9267654 | KIC 9780154 |
|---|---|---|
| $\nu_{max}$ ($\mu$Hz) | 118.62 | 216.14 |
| $M/M_\odot$ | 1.14 | 1.75 |
| $P_{rot}$ (d) | 42 | 106 |
| $P_{env}$ (d) | $\gtrsim 178$ | 214 |
| $S_{Ca\,II\,H\&K}$ | 0.199 | 0.183 |
| $S_{Mg\,I}$ | 0.571 | 0.589 |
| $S_{Ca\,II\,IR}$ | 0.638 | 0.616 |
| $S_{H\alpha}$ | 0.794 | 0.754 |

indicator, while the fraction of both LM and IM stars decreases for the Ca II IR indicator. This stands in partial contrast with the findings of Gaulme et al. (2020), who observed that the fraction of photometrically active stars indeed increases on the HB in the case of LM stars, but decreases for IM stars on the contrary. However, Gaulme et al. (2020) did not discriminate between stars on the early RGB and the medium RGB as I do in the present study. When I consider altogether LM and IM stars on the RGB for which Gaulme et al. (2020) have detected the rotation period through photometry (green and red bars and arrows in the bottom panel of Fig. 2), I also observe the same trend as Gaulme et al. (2020): the fraction of active LM stars increases on the HB compared to the RGB while the fraction of active IM stars decreases. I also note that the fraction of active stars on the HB is larger for IM stars compared to LM stars, except for the Ca II IR indicator for which the opposite trend is observed.

In addition, the right panel of Fig. 3 and Table 4 show that the increase in the fraction of active stars on the HB compared to the medium RGB is larger for IM stars compared to LM stars for almost all activity indicators, except for the Ca II IR indicator and when considering RGs that have a photometric rotation period. Overall, the fractions of active stars between the HB and the medium RGB increase by a median factor of 2.8 for IM stars and of 2.5 for LM stars when considering only chromospheric activity indicators (see Table 4). These results differ significantly from those of Gaulme et al. (2020) obtained by considering RGs that have a photometric rotation period (green and magenta cross in Fig. 3), who found that the fraction of active stars is indeed 2–3 times larger on the HB than on the RGB for LM stars, but is 1.5–2 times smaller for IM stars on the contrary. This also stands in contrast with the results I obtain by considering stars that have a photometric rotation period and are exclusively on the medium RGB, for which I find an increase of the active stars fraction on the HB by a factor of almost 5 for IM stars and of almost 15 for LM stars. This may reflect the detection bias that may affect spots in the light curves as recalled in Sect. 3.1.

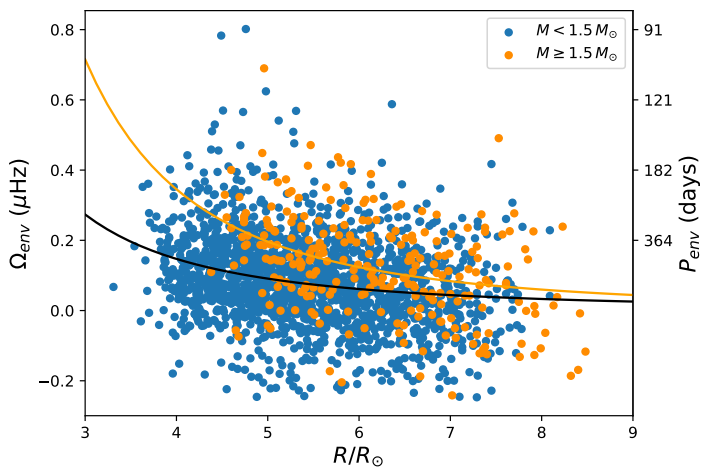

Fig. 4: Asteroseismic envelope rotation measurements from Li et al. (2024) for 1973 stars on the RGB, discriminating between LM stars (in blue) and IM stars (in orange), as a function of the stellar radius. The black and orange lines correspond to the fit of a power law to LM and IM stars, respectively.

### 3.3. Candidate red giants for planet engulfment

I note that my sample includes KIC 9267654, a LM star on the RGB that was found by Tayar et al. (2022) to have a surface rotation period from spectroscopy more than four times shorter than its envelope rotation period from asteroseismology (see Table 5), which the authors interpreted as a possible signature of planet engulfment. I also checked stars in my sample that have an asteroseismic envelope rotation rate measured by Li et al. (2024) together with a photometric surface rotation period measured by Gaulme et al. (2020) and found only one star in common, KIC 9780154, an IM star on the RGB. I found that KIC 9780154 exhibits a surface rotation twice faster than its envelope (see Table 5), which may also indicate that it has engulfed one or more planet(s). I highlight the position of KIC 9267654 and KIC 9780154 in Fig. 1.

## 4. Discussion

Here I interpret and discuss the results obtained in Sect. 3.

### 4.1. Expected evolution of magnetic activity

The efficiency of the generation of stellar magnetic fields is expected to depend on the interaction between differential rotation and sub-photospheric convection (Skumanich 1972), and is characterized by the Rossby number Ro = $P_{\rm rot}/\tau_{\rm c}$ where $P_{\rm rot}$ is the stellar rotation period and $\tau_{\rm c}$ is the convective turnover time (e.g., Noyes et al. 1984; Charbonneau 2014). I thus need to estimate how the Rossby number evolves in order to infer how magnetic activity is expected to evolve in a single-star evolution scenario.

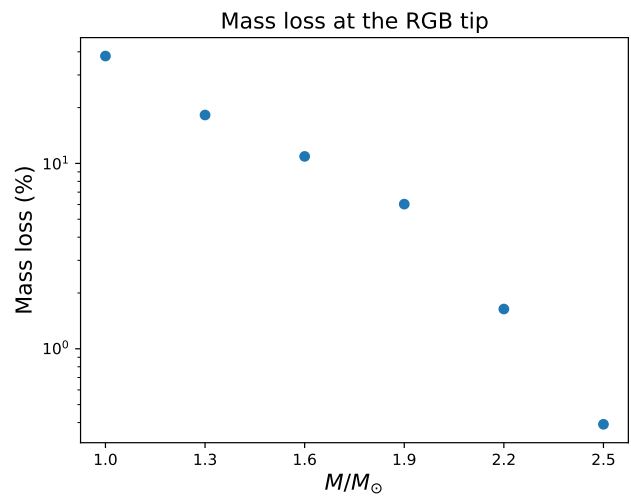

Fig. 5: Fraction of the initial mass that is lost at the RGB tip based on the Reimers law for models of different masses computed with MESA.

#### 4.1.1. Envelope rotation

I rely on existing asteroseismic measurements for stars on the RGB (Li et al. 2024). I consider the set of 1973 stars with significant non-negative envelope rotation rate measured by Li et al. (2024), and not only the 243 stars with significant positive measurements because the latter are biased toward fast rotation rates and are not fully representative of most red giants. Overall, the 1730 measurements consistent with zero are nevertheless representative of the global evolution of the envelope rotation rate for both LM and IM stars, which are expected to suffer from the same biases on their envelope rotation measurements. I infer the evolution of the envelope rotation along the RGB by fitting a power law to the envelope rotation rate as a function of the radius for LM and IM stars separately (Fig. 4), such as

$$\langle \Omega \rangle_{\rm env} = \alpha \left( \frac{R}{R_\odot} \right)^\beta. \quad (1)$$

I obtain $\alpha = 2.92$ and $\beta = -2.16$ for LM stars, while $\alpha = 11.54$ and $\beta = -2.53$ for IM stars. Therefore, the envelope rotation slows down slightly faster for IM stars than for LM stars, and steeper than an evolution in $R^{-2}$ that is expected in the frame of angular momentum conservation due to the expansion of the radius. This most probably reflects magnetic braking along the RGB, which is expected to be more efficient for faster-rotating stars (Kawaler 1988; Krishnamurthi et al. 1997; Sills et al. 2000; van Saders & Pinsonneault 2013). I also observe that the envelope rotation tends to be faster for IM stars than for LM stars on the early RGB, which is expected as magnetic braking during the main sequence is less efficient for IM stars (see Sect. 3.1). The envelope rotation period can then be derived through

$$P_{\rm env} = \frac{2\pi}{\langle \Omega \rangle_{\rm env}}. \quad (2)$$

On the HB, envelope rotation measurements are existing for only 7 IM stars (Deheuvels et al. 2015) and no LM star, which is a too small dataset to infer typical values of the mean envelope rotation for stars on the HB. However, magnetic braking and mass loss act along the RGB, removing angular momentum and thus leading to the expectation that typical envelope rotation rates would tend to be slower on the HB than on the RGB. My models indicate that mass loss remains below 2 % up to $R = 15\,R_\odot$, hence for the RGB stars in my sample, while being most important on the evolved RGB and strongly mass-dependent. My models estimate that the fraction of mass that is lost at the RGB tip with the Reimers prescription represents only $\sim 0.4$ % for a 2.5 $M_\odot$ star but up to $\sim 40$ % for a 1 $M_\odot$ star (Fig. 5). This could contribute in removing more angular momentum out of LM stars, hence slowing them down more than IM stars once they reach the HB. However, the Reimers prescription appears to be inadequate as mass loss may also be metallicity-dependent (Li 2025), hence as of today one cannot accurately predict the typical envelope rotation rates for stars on the HB, even more considering that magnetic braking also comes into play.

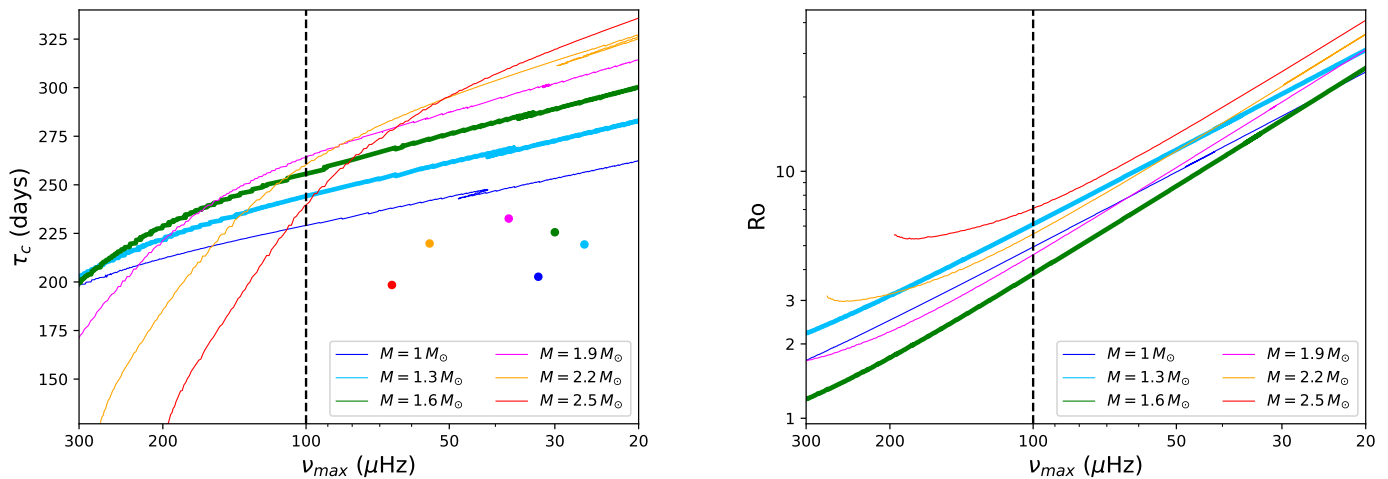

Fig. 6: Evolution of stellar properties for models of masses 1 $M_\odot$ (dark blue), 1.3 $M_\odot$ (light blue), 1.6 $M_\odot$ (green), 1.9 $M_\odot$ (magenta), 2.2 $M_\odot$ (orange) and 2.5 $M_\odot$ (red) computed with MESA, as a function of $\nu_{\max}$ (in log scale, inverted axis), from $\nu_{\max} = 300\,\mu$Hz up to $\nu_{\max} = 20\,\mu$Hz. Plain lines represent the evolution along the RGB while dots indicate median values on the HB. Models with masses of 1.3 $M_\odot$ and 1.6 $M_\odot$, representative of the median masses for LM and IM stars on the RGB, respectively, are highlighted in bold. The vertical dashed lines indicate $\nu_{\max} = 100\,\mu$Hz. *Left*: Convective turnover timescale. *Right*: Rossby number.

4.1.2. Convective turnover timescale

I used the stellar evolutionary code MESA (Paxton et al. 2011, 2013, 2015, 2018, 2019; Jermyn et al. 2023) to produce a set of models with $M = \{1.0, 1.3, 1.6, 1.9, 2.2, 2.5\}\,M_\odot$. The abundances mixture follows Grevesse & Noels (1993) and I chose a metallicity close to the solar one (Z=0.02, Y=0.28). Convection is described with the mixing length theory (Böhm-Vitense 1958) as presented by Cox & Giuli (1968), using a mixing length parameter $\alpha_{\rm MLT} = 2$. I use the OPAL 2005 equation of state (Rogers & Nayfonov 2002) and the OPAL opacities (Iglesias & Rogers 1996), complemented by the Ferguson et al. (2005) opacities at low temperatures. The nuclear reaction rates come from the NACRE compilation (Angulo et al. 1999). The surface boundary conditions are based on the classical Eddington gray $T - \tau$ relationship. Since I only aim at sketching out general features, the effect of elements' diffusion and convective core overshooting are ignored. I also include the Reimers law for mass loss along the RGB resulting from the decrease of the gravitational binding energy of the envelope (Reimers 1975). The models are evolved from the pre-main sequence to the end of the HB phase, when the center helium mass fraction drops to $10^{-4}$. The end of the main sequence is identified as the model cor-responding to a center hydrogen mass fraction of $10^{-3}$. I identify the transition between the subgiant branch and the RGB as the moment when the helium core mass reaches the Schönberg–Chandrasekhar limit, i.e. 10 % of the stellar mass (Schönberg & Chandrasekhar 1942). I estimate the convective turnover time based on the analytical approach of Corsaro et al. (2021). They showed that the convective flux can be approximated by

$$F_{\rm c} \simeq \frac{15}{8\pi\sqrt{2}}\,\alpha_{\rm MLT}^2\,\frac{M}{R^3}\left(\frac{GM}{R}\right)^{3/2}(\nabla - \nabla_{\rm ad})^{3/2}\,, \quad (3)$$

where $\nabla$ is the temperature gradient and $\nabla_{\rm ad}$ is the adiabatic temperature gradient. Since $F_{\rm c} \simeq L/R^2$, one can obtain that

$$(\nabla - \nabla_{\rm ad}) \simeq \left(\frac{8\pi\sqrt{2}}{15}\right)^{2/3}\alpha_{\rm MLT}^{-4/3}\left(\frac{LR}{M}\right)^{2/3}\frac{R}{GM}. \quad (4)$$

The convective velocity can then be estimated through

$$v_{\rm c} \simeq c_{\rm s}\sqrt{\nabla - \nabla_{\rm ad}} \simeq \left(\frac{8\pi\sqrt{2}}{15}\right)^{1/3}\alpha_{\rm MLT}^{-2/3}\left(\frac{LR}{M}\right)^{1/3}, \quad (5)$$

where $c_{\rm s} \simeq \sqrt{GM/R}$ is the sound speed. The convective turnover time $\tau_{\rm c}$ finally corresponds to the ratio between the extent of the convective envelope and the convective velocity, hence

$$\tau_{\rm c} \simeq (R - r_{\rm BCZ})\left(\frac{15}{8\pi\sqrt{2}}\right)^{1/3}\alpha_{\rm MLT}^{2/3}\left(\frac{M}{LR}\right)^{1/3}, \quad (6)$$

where $r_{\rm BCZ}$ is the radius coordinate associated with the base of the convective envelope, and $\alpha_{\rm MLT} = 2$. The resulting $\tau_{\rm c}$ are represented in the left panel of Fig. 6. I focus in the following on models with 1.3 $M_\odot$ and 1.6 $M_\odot$ (highlighted in bold) because these masses are very close to the median masses of LM and IM stars in my sample, that are 1.28 $M_\odot$ and 1.65 $M_\odot$, respectively. The convective turnover time is systematically shorter for LM stars than for IM stars along the RGB, implying that LM stars require a faster envelope rotation to reach a similar Rossby number and hence, a similar level of magnetic activity, than IM stars.

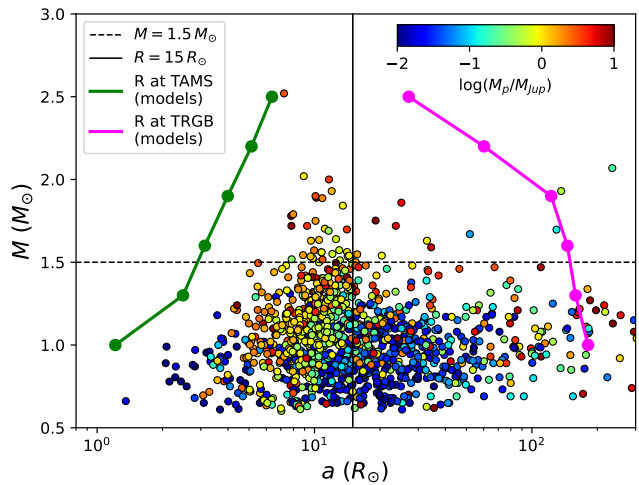

Fig. 7: Properties of the exoplanets with known mass around stars with $0.6 \leq M \leq 2.5\,M_\odot$ and $R \geq 4 \leq R\,R_\odot$. Mass of the host star as a function of the semi-major axis of the planet. The color code represents the decimal logarithm of the mass of the planet, in Jupiter masses. The horizontal dashed lines indicate a mass of $M = 1.5\,M_\odot$. The green and magenta lines and dots indicate the radius at the end of the main sequence (TAMS) and the tip of the red giant branch (TRGB), respectively, for models of $M = \{1, 1.3, 1.6, 1.9, 2.2, 2.5\}\,M_\odot$. The vertical continuous line indicates a radius of $R = 15\,R_\odot$.

Additionally, $\tau_c$ is systematically shorter on the HB compared to the RGB, implying that a faster envelope rotation is required on the HB to reach a similar Rossby number and hence, a similar level of magnetic activity, than on the RGB.

### 4.1.3. Rossby number

I then estimate the Rossby number based on the measured envelope rotation periods from Sect. 4.1.1 and the modelled convective turnover time from Sect. 4.1.2, which is represented in the right panel of Fig. 6 for stars of different masses. As noted in Sect. 4.1.1, we have envelope rotation measurements for a too small number of stars on the HB to infer typical values of the mean envelope for such stars, hence to infer how the Rossby number behaves on the HB.

All along the RGB, the Rossby number is systematically smaller for IM stars compared to LM stars, hence we expect IM stars to be more active than LM stars all along the RGB. However, this is not what I observe, as I find in Sect. 3.1 that the fraction of active LM stars tends to become larger than that of IM stars along the RGB. Moreover, we expect magnetic activity to decrease along the RGB since the Rossby number increases, while I find on the contrary that the fraction of active LM stars tends to increase along the RGB. The results I obtain in Sect. 3.1 are therefore not expected from single star evolution for LM stars, and could potentially result from the engulfment of planet(s), hypothesis that I explore below.

### 4.2. Properties of known planets around main-sequence stars

I check here the distribution of the 6065 known planets, in particular their distance to their host star as a function of their mass and of the mass of their host star, using the Extrasolar Planets Encyclopaedia[1]. To match the sample of RGs used in this study (Gaulme et al. 2020; Gehan et al. 2022, 2024), I focus on planets around stars with $0.6 \leq M \leq 2.5\,M_\odot$. In order to have an overview of planets around main-sequence stars, I focus in addition on the planets around stars with $R \leq 4\,R_\odot$. I am left with 3007 planets, among which 1391 with known mass.

I strikingly observe that there are significantly less planets known around IM stars than LM stars (Fig. 7). This could reflect an observational bias, as a large number of planet studies focus on LM stars, and as planet detection through both the radial velocity and transit techniques can be more challenging around IM stars that are bigger, hotter, and faster rotators, therefore more magnetically active, than LM stars. However, theory under the core-accretion planet formation scenario supports a global dearth of planets around IM stars (Laughlin et al. 2004; Ida et al. 2013). On the one hand, stars with $M \sim 1.3$–$2.1\,M_\odot$ have been observed to host twice as many massive planets as Sun-like stars (Johnson et al. 2010; Vigan et al. 2012), as reflected in Fig. 7 where I no longer see low-mass planets around stars with $M \gtrsim 1.3\,M_\odot$. This comes from the fact that more massive stars are expected to have more massive disks during the pre-main sequence (Kennedy & Kenyon 2008). On the other hand, the timescale available for giant planets to form, hence the number of formed planets, is largely determined by the lifetime of the protoplanetary disc (Ronco et al. 2017; Guilera et al. 2020; Venturini & Helled 2020; Venturini et al. 2020), which decreases with stellar mass (Ribas et al. 2015; Kunitomo et al. 2021; Komaki et al. 2021; Ronco et al. 2024). Moreover, the core-accretion scenario further supports a dearth of close-in planets around IM stars. Indeed, planets are expected to form further out around IM stars than around LM stars because the various snow lines and photoevaporation limits are more distant around more luminous and therefore more massive stars (e.g., Murillo et al. 2022; Giacalone & Dressing 2025). Hence, the dearth of planets observed around IM stars in Fig. 7 is likely to be real, although the observed population of planets around IM stars may suffer from observational biases and their number may be underestimated since more distant planets are harder to detect by both the radial velocity and transit methods. I also observe that a significant fraction of the known planets are orbiting at a distance shorter than $15\,R_\odot$, which is the upper radius for the red giants in my sample. Hence, LM stars in my RGB sample, which are not very evolved on the RGB, are nevertheless already much more susceptible of engulfing a planet than IM stars as they evolve along the RGB.

It is worth noticing that the amount of angular momentum that is deposited in the envelope of the host star does not only depend on the number of engulfed planets, but also on the mass of the engulfed planet(s) and of the star as well as on the initial distance between the star and the planet. Faster rotation rates are reached for massive planets, low-mass stars, and smaller orbital distances (Privitera et al. 2016b). The last condition may seem counterintuitive, however it derives from the fact that the shorter the orbital distance, the earlier the engulfment event on the RGB that corresponds to a smaller moment of inertia of the stellar envelope, and hence the more important the acceleration of the stellar rotation (Privitera et al. 2016b). Figure 7 shows that LM not only host low-mass planets but also massive planets within short orbital distances. All these considerations are compatible with the milder decrease or even increase of the active LM star fraction along the RGB that I find in Sect. 3.1.

---

[1] exoplanet.eu/catalog on September 3, 2025

On the HB, one might expect to observe a fraction of active stars that is larger for LM stars than for IM stars, assuming LM stars keep engulfing more planets than IM stars on the upper RGB, however I observe the contrary for almost all the activity indicators used in this study (Fig. 2 and Table 4). As mentioned in Sect. 4.1.1, the situation is however not straightforward because LM stars lose much more mass on the upper RGB than IM stars, which may contribute to making them less active than IM stars once on the HB.

I note that my results do not point toward engulfment of a stellar companion rather than that of a planet. Indeed, the fraction of multiple systems increases with stellar mass (Whitworth & Lomax 2015; Moe & Di Stefano 2017), while my results indicate that LM red giants engulf more companions. Moreover, stellar mergers generally result in a larger increase of the mass of the resulting star than in the case of planet engulfment. Hence, LM stars undergoing mergers would become IM stars, making IM stars more plausible to be the result of stellar mergers than LM stars. In this scenario, an activity increase associated to mergers would a priori result in more active IM stars compared to LM stars. However, I observe the opposite behaviour, which further favours planet engulfment over stellar mergers as an explanation of increased activity in LM stars compared to IM stars.

## 5. Conclusions

Gaulme et al. (2020) observed that the fraction of photometrically active RGs, i.e. exhibiting rotational modulation in their light curve, is larger on the HB than during the previous RGB phase for LM stars, while they observed the opposite behaviour for IM stars. They interpreted such observation as a potential signature of planet engulfment, since LM stars appear more susceptible of engulfing planets as they reach a much larger radius at the RGB tip than IM stars. However, spot detection through photometry can suffer from important biases, which could potentially impact the fractions of active stars computed by Gaulme et al. (2020).

In order to get a more complete view of the activity of RGs with different evolutionary stages and masses, I use several spectral lines whose depth trace the level of chromospheric activity indenpently of the detection of photospheric spots, namely the Ca II H&K lines (3969 & 3934 Å), one of the Mg I lines (5184 Å), the H$\alpha$ line (6563 Å) and one of the Ca II IR lines (8542 Å) that were studied by Gehan et al. (2022, 2024) for ∼ 3000 RGs with $4 \leq R \leq 15 R_\odot$ using Data Release 7 of LAMOST. I computed the fractions of active LM and IM stars at two different evolution stages on the RGB as well as on the HB, using $\nu_{max}$ as a proxy of evolution along the RGB for the activity indicators mentioned above.

I observe that the fraction of active stars shows different trends for LM and IM stars along the RGB, decreasing by a median factor of 2.8 for IM stars while increasing by a median factor of 1.1 for LM stars. Such an increase is not expected from the evolution of the Rossby number for single stars given by asteroseismic measurements of the envelope rotation and models of the convective turnover time computed with MESA, and is instead compatible with a scenario of planet engulfment. This is reinforced by the observed dearth of planets observed around IM main-sequence stars that is likely to be real, as predicted by the theory of planet formation. This makes LM stars much more susceptible of engulfing planets than IM stars as they evolve along the RGB. The properties of the known planets around main-sequence stars moreover indicate that the effects of planet engulfment on the evolution of magnetic activity of LM versus IM stars are inded expected to be significant already for $R \leq 15 R_\odot$ on the RGB, hence for stars on the RGB considered in this study.

Between the medium RGB and the HB, the fraction of active stars tends to increase for both LM and IM stars, by a median factor of 2.5 for LM stars and of 2.8 for IM stars. This stands in partial contrast with the findings of Gaulme et al. (2020), who instead observed that the fraction of photometrically active stars decreases on the HB for IM stars, and can be explained by the fact that Gaulme et al. (2020) did not discriminate between stars on the early RGB and the medium RGB as I do in the present study. The evolution of the active star fraction between the medium RGB and the HB is once again not expected from single star evolution because magnetic braking and mass loss are expected to slow down the envelope rotation on the one side, and the convective turnover timescales are shorter compared to the RGB on the other side, leading to the expectation of a decrease of magnetic activity between the medium RGB and the HB. The active star fraction increase that I observe between the medium RGB and the HB is compatible with planets keeping being engulfed on the evolved RGB. I note that the fraction of active stars on the HB tends to be larger for IM stars compared to LM stars overall, which is intuitively not expected in the frame of planet engulfment mainly by LM stars, however the situation is more complicated to interpret on the HB where mass loss during the evolved RGB phase is expected to decrease magnetic activity and is significantly more important for LM stars compared to IM stars, making this observation not necessarily incompatible with a scenario of planet engulfment.

Finally, I discover that the IM RGB star KIC 9780154 may have engulfed one or more planet(s) as its surface rotation from photometry (Gaulme et al. 2020) is twice faster than its envelope rotation from asteroseismology (Li et al. 2024). Characterizing planet engulfment by RGs provides insights on the evolution of planetary systems and on the fate of planets around white dwarfs, which represent the ultimate evolutionary stage of ∼ 97 % stars.

*Acknowledgements.* I was supported by a postdoctoral fellowship from the CNES. I dedicate this work to the late Patrick Gaulme, an exceptional and much-missed collaborator without whom it would never have seen the light of day. I thank the referee, who graciously revealed his identity, for his helpful and insightful comments, which greatly contributed to improving this manuscript.

## Appendix A: Impact of using a 3-$\sigma$ prediction interval instead of 2-$\sigma$

I check to which extent the computed active star fractions and the overall trends obtained in this study are impacted by taking the 3-$\sigma$ prediction interval around the fit in Fig. A.1 instead of the 2-$\sigma$ prediction interval I use in Fig. 1. The recomputed active star fractions are indicated in Table A.1 and shown in Fig. A.2, highlighting that the overall trends remain unchanged. The active star fraction keeps decreasing for IM stars and increasing for LM stars along the RGB, at the exception of the H$\alpha$ indicator for which the active fraction increases for IM stars and remains constant for LM stars. The active star fraction on the low RGB remains higher for IM stars than for LM stars, once again at the exception of the H$\alpha$ indicator. Additionally, the fraction of active stars increases for both LM and IM stars on the HB compared to the medium RGB, except for the Mg I indicator for which it decreases for LM stars and for the Ca II IR indicator for which it remains at zero. The main change compared to using the 2-$\sigma$ prediction interval is that the active star fraction on the HB is now larger for LM stars compared to IM stars, at the exception of the H$\alpha$ indicator.

Using the the 3-$\sigma$ prediction interval instead of 2-$\sigma$ significantly decreases the number of stars that are considered active and thus the derived active star fractions, which are now much lower and even fall at zero in several cases. They do not agree overall with the active star fractions derived by Gaulme et al. (2020) from the photometric detection of rotation periods.

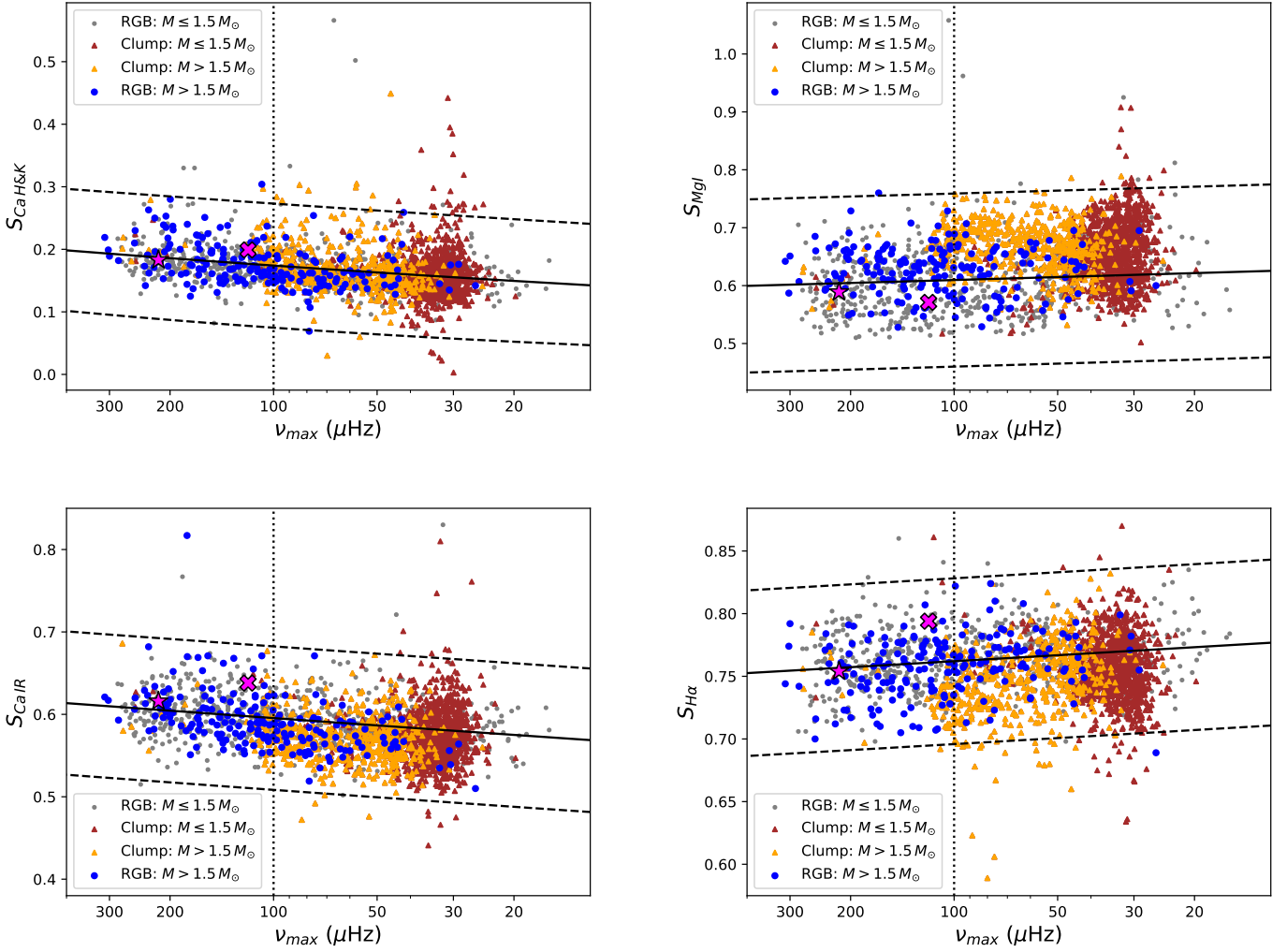

Fig. A.1: Same as Fig. 1, except that the dashed lines represent the 3-$\sigma$ prediction interval around the fit.

Table A.1: Characteristics of the active sample.

|  | RGB, active | | | | HB, active | |
|---|---|---|---|---|---|---|
|  | $M \leq 1.5\,M_\odot$, $\nu_{max} \geq 100\,\mu Hz$ | $M \leq 1.5\,M_\odot$, $\nu_{max} < 100\,\mu Hz$ | $M > 1.5\,M_\odot$, $\nu_{max} \geq 100\,\mu Hz$ | $M > 1.5\,M_\odot$, $\nu_{max} < 100\,\mu Hz$ | $M \leq 1.5\,M_\odot$ | $M > 1.5\,M_\odot$ |
| Ca II H&K | 0.74 % (2) | 1.43 % (4) | 0.90 % (1) | 0 % (0) | 2.12 % (21) | 1.97 % (10) |
| Mg I | 0.36 % (1) | 2.13 % (6) | 0.89 % (1) | 0 % (0) | 1.70 % (14) | 0.57 % (2) |
| H$\alpha$ | 0.35 % (1) | 0.34 % (1) | 0 % (0) | 1.32 % (1) | 1.54 % (16) | 4.17 % (23) |
| Ca II IR | 0.35 % (1) | 0.68 % (2) | 0.88 % (1) | 0 % (0) | 0.77 % (8) | 0 % (0) |

**Notes.** Same as Table 2 for chromospheric indicators, except that the threshold to consider that stars are active is now the 3-$\sigma$ prediction interval around the fit in Fig. A.1.

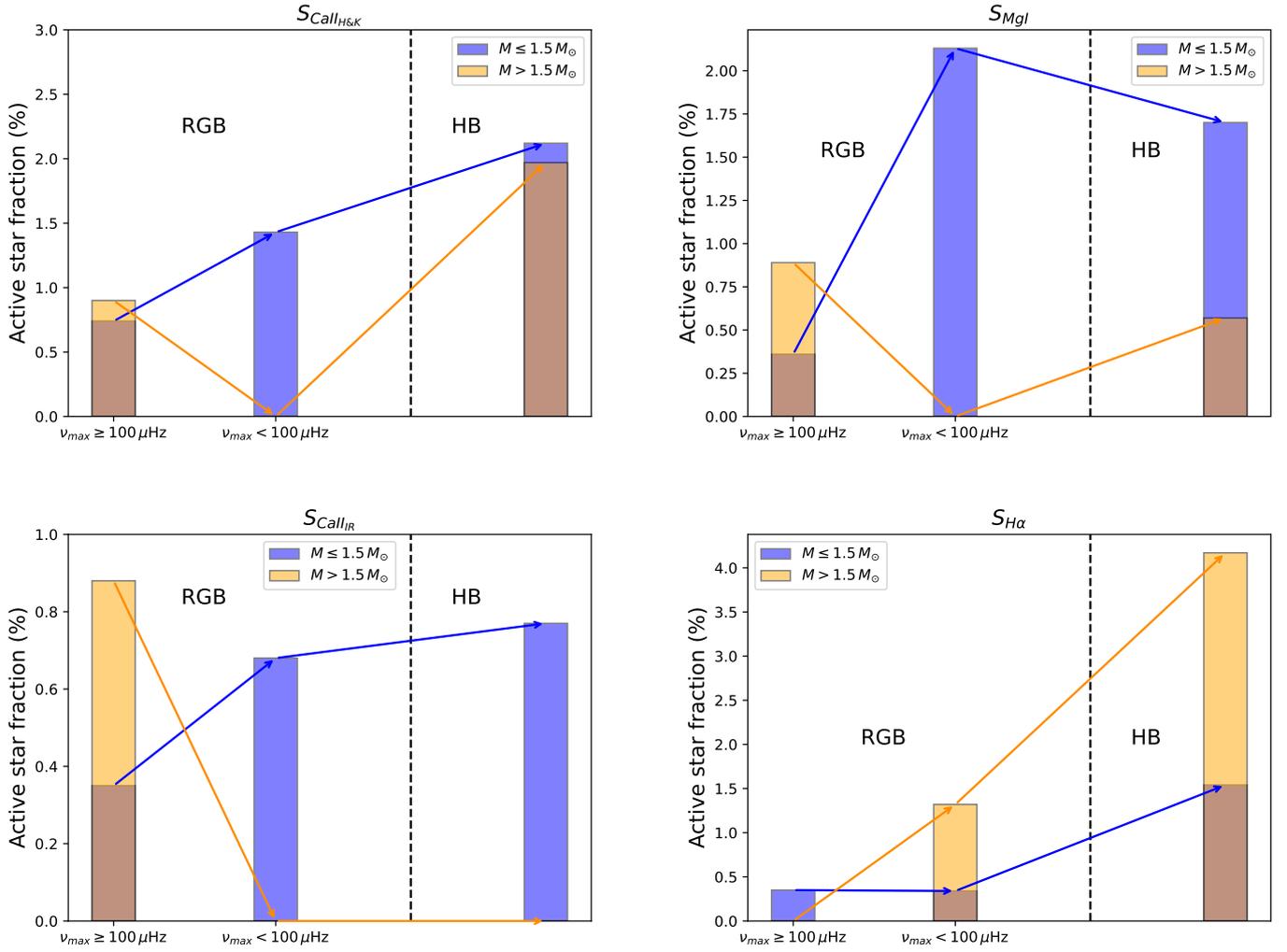

Fig. A.2: Same as Fig. 2 for chromospheric indicators, except that the active fractions are determined using the 3-$\sigma$ prediction interval around the fit in Fig. A.1.